# CROSS PLATFORM APP
# A COMPARATIVE STUDY

Paulo R. M. de Andrade, Adriano B. Albuquerque

Postgraduate program in applied information University of Fortaleza - UNIFOR
Fortaleza - CE, Brazil

Otávio F. Frota, Robson V Silveira, Fátima A. da Silva
College Estacio FIC of Ceara Fortaleza - CE, Brazil

## ABSTRACT

*The use of mobile applications is now so common that users now expect companies whose services which they consume already have an application to provide these services or a mobile version of your site, but this is not always simple to do or cheap. Thus, the hybrid development has emerged as a potential alternative to this need. The evolution of this new paradigm has taken the attention of researchers and companies as viable alternative to the mobile development. This paper shows how hybrid development can be an alternative for companies provide their services with a low investment and still offer a great service to their clients.*

## 1. INTRODUCTION

Because of the exponential increase in the need for people to stay connected with everything and everyone through the internet in search of information and communication, increases the need for suitable sites and applications to this new reality. According to the International Data Group [5], since 2010, the amount of mobile devices has increased more than 15% per quarter, which further emphasizes the need for adequacy of companies, government agencies and universities to this new reality. For companies, the lack of technical knowledge has been the obstacle for not implement good ideas due to costs necessary for training. For this reason, begins to be a need to support a range of popular platforms such as iOS, Android, WP8, Blackberry, etc.

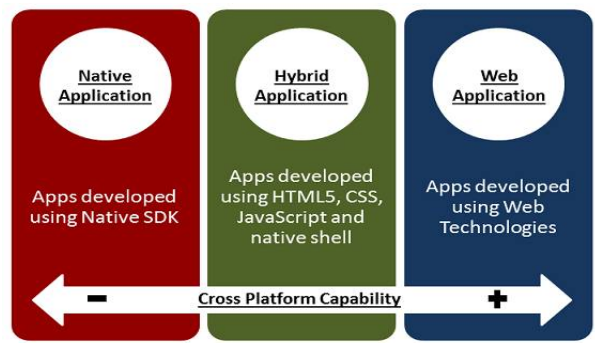

The development of native applications require a high level of specialized knowledge in programming. For example, the development of native applications for iPhone requires knowledge of Objective-C and when we talk about Android, you need knowledge of JAVA programming language, even more commonly, still requires specific training for mobile development. A multiplatform architecture would be a solution to make this difficult task into something much more affordable, and with the possibility of the development using Web methods or hybrid application development.

The development using HTML5, CSS3 and JavaScript allows a single application for smart phones to work on multiple operating systems using the same markup language of websites and requires a minimal level of investment in technical knowledge and time. The frameworks (known as "shells") minimize the need for specialized programming language and increases the power of use of native application APIs [8].

This paper address the architectures for web development and needs for multiplatform development. It presents the PhoneGap and the Intel App Framework. The following section presents a case study using the two frameworks and the results of empirical tests. Finally, the last section presents the conclusion and suggestion for future work.

## I. EASE OF USE

In order to understand the multiplatform architecture that we first understand the types of strategies used in the development of applications for mobile devices. Figure 1 shows the comparison between the strategies of application development and indicates its multiplatform, increasing fitness from left to right [7].

Fig. 1. Strategies for mobile development

Hybrid applications combine the advantages of both types of the development of applications (web and native) and the best choice is to create cross-platform applications. The main advantages of using a hybrid technology are:

- Multiplatform & Sharing code: coding once and use it to make the "deploy" on various mobile platforms. Uses the same code "UI" interface for multiple devices;
- Make native calls to hardware using the "Native Shell" through of the JavaScript;
- Offline mode allows access to the same applications that the internet is not readily available.
- Allows a large number of users can get the application due to its multiplatform nature.
- One can avail the mobile processing, which is not possible with web applications.
- Distribution through official stores each platform transmits a native feeling regarding updates.

The web applications also allow a multiplatform fitness, but in a more limited way. Developers will not be able to take advantage of the device hardware and native UI. Furthermore, the local processing means of a browser will not be a good choice for applications whose performance is critical. In case the performance is not a problem and just want to host a web application then this kind of application is sufficient, eliminating the use of a "third party Framework" and further promote the use of existing knowledge of web technology.

Native applications offer advantages multiplatform very limited or even nonexistent. For this reason is that they are not advisable for a multiplatform architecture [3]. Figure 2 illustrates the main pros and cons of each type of application.

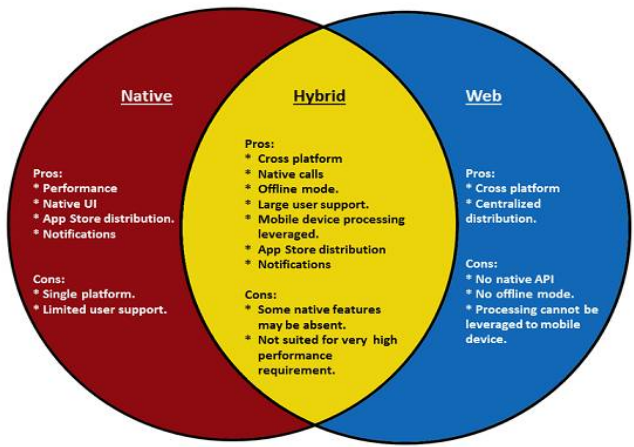

Fig 2. Pros and cons of each type

In a multi-platform architecture using hybrid methods, the development of an application uses Web technologies such as HTML5, JavaScript and CSS but that run inside the "Native Shell" of the Mobile platform. Thanks to the increasing sophistication of multi-platform tools, performance has improved dramatically, and both the look and feeling that we have to use the application is almost as good as the native UI.

## 2. PHONEGAP

The PhoneGap is a framework that has the function of "Native Shell". It created by Nitobi in 2008 as an open source solution for building cross-platform mobile application, passing the first to support the iPhone, Android and Blackberry 4 later in the Symbian and WebOS and Windows Phone 7 finally in 2011, Adobe acquires Nitobi Software. In October 2011, PhoneGap has donated to the Apache Software Foundation (ASF), under the name Apache Cordova. Through the ASF, the development of PhoneGap passes to ensure open government project. Remaining free and open source under the Apache License Version 2.0 [2].

The PhoneGap uses the methodology write-once, where the idea is to write the code only once and by importing, exporting it as a native application. Currently supports all major platforms such as iOS, Android, Blackberry 10, Windows Mobile, Windows Phone (7 and 8), Firefox OS, Ubuntu and Tizen. Through its library and the native code generated, PhoneGap allows API calls for smart phone hardware [2].

To understand better, PhoneGap is a set of APIs that allows the developer to access the native device functions like camera, calendar, GPS and other by JavaScript, HTML5 and CSS3, instead of device-specific languages such as Objective-C. The development is like any site and therefore offers greater ease of learning. Dispensing, for example, long hours of dedication to more complicated systems, such as Java and reduce project costs [6]. Figure 3 we can see all the APIs supported by the framework.

| | android | blackberry10 | Firefox OS | ios | Ubuntu | Windows Phone |
|---|---|---|---|---|---|---|
| cordova CLI | ✔ Mac, Windows, Linux | ✔ Mac, Windows | ✔ Mac, Windows, Linux | ✔ Mac | ✔ Ubuntu | ✔ Windows |
| Embedded WebView | ✔ (see details) | ✘ | ✘ | ✔ (see details) | ✔ | ✘ |
| Plug-in Interface | ✔ (see details) | ✔ (see details) | ✘ | ✔ (see details) | ✔ | ✔ (see details) |
| **Platform APIs** | | | | | | |
| Accelerometer* | ✔ | ✔ | ✔ | ✔ | ✔ | ✔ |
| Camera* | ✔ | ✔ | ✔ | ✔ | ✔ | ✔ |
| Capture* | ✔ | ✔ | ✘ | ✔ | ✔ | ✔ |
| Compass* | ✔ | ✔ | ✘ | ✔ (3GS+) | ✔ | ✔ |
| Connection* | ✔ | ✔ | ✘ | ✔ | ✔ | ✔ |
| Contacts* | ✔ | ✔ | ✔ | ✔ | ✔ | ✔ |
| Device* | ✔ | ✔ | ✔ | ✔ | ✔ | ✔ |
| Events | ✔ | ✔ | ✔ | ✔ | ✔ | ✔ |
| File* | ✔ | ✔ | ✘ | ✔ | ✔ | ✔ |
| Geolocation* | ✔ | ✔ | ✔ | ✔ | ✔ | ✔ |
| Globalization* | ✔ | ✘ | ✘ | ✔ | ✔ | ✔ |
| InAppBrowser* | ✔ | ✔ | ✔ | ✔ | ✔ | ✔ |
| Media* | ✔ | ✔ | ✘ | ✔ | ✔ | ✔ |
| Notification* | ✔ | ✔ | ✘ | ✔ | ✔ | ✔ |
| Splashscreen* | ✔ | ✔ | ✘ | ✔ | ✔ | ✔ |
| Storage | ✔ | ✔ | ✘ | ✔ | ✔ | ✔ localStorage & indexedDB |

Fig. 2. PhoneGap API's

Fig. 3.

The creation of a generic and offline using HTML, CSS and JavaScript application and its testing in several screen sizes, test the use of imported APIs in its development (as Geolocation, Camera and Notification APIs) is possible using an emulator for Google Chrome called "Ripple Emulator". After creating the application, it is possible in PhoneGap v.3.0+ create projects for each platform through command line using "Node.js". When creating a project, the generated files are accessible natively in each program to perform the necessary customizations for each platform and test on emulators.

A particularity is found that for iOS apps that you need use a MAC for continue you development, because the generated project should be opened with xCode and requires its own libraries as well the project for Windows Mobile needs be done using Windows 8.0 or higher. Projects for Android, Blackberry OS and Firefox OS work on any platform. An alternative is to use PhoneGap Build Online and perform the conversion in native code independent of the operating system being used [2].

## 3. INTEL APP FRAMEWORK

The Intel Company may still be known for creating the chips that power the world's desktops, but the company has been trying to expand into the world of software for many years. The Intel App is one of their latest designs. The Intel App Framework is a framework for building cross-platform mobile applications using HTML5 technologies. He came when based jqMobi, a mobile optimized version of jQuery, which was created by the team behind appMobi. Intel acquired the tools and personnel jqMobi in February 2013 [4].

The Intel App Framework is a free and open source under an MIT license. Combined with its lightweight JavaScript library, it provides a basic MVC structure and various components of the user interface. Instead of imitating the appearance and behavior native, he chose to offer their own style for use on all platforms. Over time, styles added to mimic the display of native apps, at the request of users, being free to choose the style to use, and maybe carry a different style according to the operating system of the mobile device where the application is running. We can also customize the styles (css and html) using the "Style Builder" tool. The Intel App Framework can be used together easily with PhoneGap to create native applications instead of web applications, which is their standard.

One of the most notable differences is a smaller library to manipulate the DOM, which offers the most important features of jQuery without the slower functions. Intel claims that its library is faster and more robust response to the mobile internet. The functionality and the structure is almost the same as jQuery, and has a perfect compatibility with it, allowing you to install a plug-in jQuery [4].

The best part may be the largest collection of tools that includes a website and a Java-based build and test their applications client. There is also a Windows executable for those developers who want everything running locally. The entire structure is marked in HTML, and JavaScript does the job. In the Intel Style Builder tool, just choosing the colors, it generate a new CSS with the style for your application. The Framework keeps some of the elements used less, like the carousel, as separate plug-ins that you can add when you need it. This keeps the resulting code smaller and therefore easier to deploy. This framework has its own native APIs to call the hardware but usually use the phonegap together to use the hardware, so, the Intel App Framework just assuming the role of HTML framework.

Is also possible to use the Intel App Starter, an online tool, for help you create an initial project. You can drag and drop DIVs and components in the right place and test the results in your browser. You will create all the basic structure of your application, such as pages, menus, choose transitions, simpler forms and customizations header. Upon completion of construction, it is possible to download the generated html code and use it to continue the application [4].

## 4. CASE STUDY AND RESULTS

The Cagece (a company of basic sanitation Brazilian public sector) noted the need to create a new management practice for the creation and implementation of a new way of managing the problems reported by the population and for simplifying the process. After a benchmarking from the industry and with a vision of the need for portability and usability, the Cagece decided that to obtain this new practice should be through a mobile application. The board of the company approved the presented project created. They used the Agile PDD methodology [1] to plan and manage this project.

The app was initially an idea where the population would be able to, at any time, inform about problems encountered of leakage of water and sewage, frauds and other general occurrences.
The major problem was the lack of skilled labor within the company to develop for iOS. Initially it was thought to be held in an internal training tool, however only one application did not justify the investment and have a small learning curve. Another option was to perform the outsourcing for development through a software factory.

Within the company, there was a web development team, which proposed to study the possibilities and present to the business area, proposals to make the project viable. After a few weeks of study and comparisons of available technologies and the need for each, the team chosen a hybrid application development using HTML 5 as the basis of the application, PhoneGap v.3.1 as "Native Shell" and the Intel App Framework v2.0 as the basis of the interface. To integrate with the company trading system was used a JBoss Web Service.

The client when open the application on your smart phone or tablet will have a unified access (the same information on web site is used in mobile) where he could consult his registration, last 12 invoices data, open invoices (can view or request a duplicate by the e-mail), consumption history and locate a service store nearest. Furthermore, he can register his occurrence of water shortage, water leaks, sewage leak, fraud, services not completed, water holes and sewage or other events. In the occurrence he can give her address automatically by the GPS position and he has the option to send a picture related using the camera or chosen from the gallery.

In the application, the client can also access the news that are on the company's website, he can find different tips (the system are some tips on avoiding leaks, cleaning of water tanks, water and sewage treatment), list of services and their prices, frequently asked questions from individuals and companies, and direct contact with the company channels.

The entire development of the hybrid application lasted 12 weeks. It was the first application developed by the team. For make a comparison, in parallel, the Java development team, with some experience in mobile development, has also developed a version of the same application, taking 10 weeks to complete.

After finishing the two developments, were selected 60 employees of the company, experienced in the use of Android devices, to use the application. These devices were chosen because they are known, in some cases, to have a more limited performance. Were selected 30 tablets Samsung GT-P3110 and 30 mobile phones Samsung S-i9000. Each employee received one of the devices. In the first round of tests, only 12 devices of each type had the native application, while the others had the hybrid application. The table 1 illustrates the division.

Table 1. The division of devices in the round one

| Devices | Native App | Hybrid App |
|---------|-----------|-----------|
| GT-P3110 | 12 | 18 |
| S-i9000 | 12 | 18 |

The instructions for the employees were to use the application for 2 weeks and give your feedback on the experience. After use, half the amount of devices were reversed, but was told to all employees that they would receive a device with a different version of the test first, applying the placebo effect. Employees would use two more weeks to give a new feedback. The table 2 illustrates the new division of devices.

Table 2. The division of devices in the round two

| Devices | Same Native | Changed Native to Hybrid | Same Hybrid | Changed Native to Hybrid |
|---------|-------------|--------------------------|-------------|--------------------------|
| GT-P3110 | 4 | 8 | 10 | 8 |
| S-i9000 | 5 | 7 | 11 | 7 |

In the second round of testing, 76.67% (46 of 60) of the users reported that they have not noticed a difference in the use of the application, stating that had the same experience. The other 23.33% users reported that they perceived that an application was relatively faster over another in their response to use with the native applications. On the other hand, the users who had not changed the tablets, only 10% (6 of 9) of the users had the impression its initial implementation used was faster than the second application, even if they were the same. As the result, we can see that only 13.33% (8 of 60) users realized the difference in performance between the native and hybrid apps.

This comparison helps to see that the hybrid application, even taking a more time in this case, is a viable alternative for companies when the need is to create applications for different platforms. The performance of the current frameworks has increasingly approached the native applications. Another positive point is a shorter learning curve for new applications by taking advantage the common knowledge of the most WEB developers, not making it necessary for learning a new language. This comparison also shows that native development is a viable option only when a deeper use of specific APIs, frameworks and not supported by harnessing the full power of the displays and UIs is required.

Today, the company distributes the mobile app created (Cagece App) for Android and iOS (smart phones and tablets) and gets as a result a large participation of the population with respect to ability to inform places of water and sewers leaks using it. Since the launch in August 2013 until the day December 31, 2014, the users opened 3,188 occurrences through the mobile devices. The application had over 12,000 downloads in this period and has received good reviews from users.

Other data analyzed are those relating to the attendance emission the new copy of the clients' bill in service stores. During the period April-December 2013 were generated 253,966 in service stores throughout the state while in 2014 were 423,727 for the same period. In 2014, the app realized 234,873, which represents 55.43% of the total. This is an excellent number that represents the number of not printed papers (that's contributes to reduce emission of $CO_2$) and the virtual attendance, which contributes reducing the waiting time of customers and lines in the stores.

## 5. CONCLUSION

Native applications can provide a good user experience, but maybe the companies had no money or expertise to develop natively. A hybrid approach offers a simple solution for developing applications for smart phones and tablets. Write the code once and deploy to different operational systems will help companies to quickly launch their mobile applications and reduce maintenance costs. Hybrid structures are suitable options for the real benefits in the use of applications for business or for education.

Are suggested as future work more detailed metrics on the learning curve, performance and savings generated for hybrid applications and different sizes and needs.